\documentclass[runningheads]{llncs}
\usepackage[T1]{fontenc}
\usepackage{amsmath}
\usepackage{float}
\usepackage{booktabs}
\usepackage{makecell}
\usepackage{url}
\usepackage{hyperref}
\usepackage{graphicx}
\begin{document}
\title{Improving Service Availability in KubeEdge-Based Architectures Using Lightweight Intrusion Detection}

\titlerunning{Improving Service Availability in KubeEdge Using Lightweight IDS}  
%
%

\author{
Harrol Ndjeudji Kuibou\inst{1}\orcidID{0009-0007-4518-3137}\and
Mostafa Anouar Ghorab\inst{1}\orcidID{0000-0003-2235-7126}\thanks{Corresponding author.}\and
Mohamed Aymen Saied\inst{1}\orcidID{0000-0002-9488-645X}
}

\authorrunning{H. N. Kuibou et al.}

\institute{
Laval University, Quebec City, Quebec, Canada\\
\email{
harrol.ndjeudji-kuibou.1@ulaval.ca,
mostafa-anouar.ghorab.1@ulaval.ca,
mohamed-aymen.saied@ift.ulaval.ca
}
}

\maketitle              
\begin{abstract}
The increasing adoption of the Internet of Things (IoT) and cloud computing has accelerated the evolution of edge computing paradigms \cite{shi2016edge}. Industry forecasts estimate that the number of connected IoT devices will reach approximately 50 billion by 2030, following an estimated 38 billion connections by 2025 \cite{b30}, resulting in an unprecedented growth in data generation. This trend necessitates efficient, scalable, and secure data processing mechanisms at the network edge. Consequently, ensuring the reliable management and protection of IoT applications and devices has become a critical challenge.

In this context, KubeEdge extends cloud-native capabilities to edge environments, enabling distributed orchestration while introducing new security concerns. This paper investigates the security of container images in IoT-driven and distributed edge architectures. Specifically, we analyze the impact of major security threats, including Denial of Service (DoS) attacks and malicious container deployments, on the availability and operational stability of KubeEdge-based systems.

To address these challenges, we propose a lightweight Recommended Intrusion Detection Rule Set (RIDRS) tailored for resource-constrained edge environments. The proposed approach improves system resilience by enabling timely detection and mitigation of security threats. We define system stability as the ability to maintain consistent operational behavior and to recover autonomously under adversarial conditions. Experimental results demonstrate that RIDRS significantly reduces system downtime and enhances service availability, particularly in scenarios involving code injection and malicious pod deployment attacks.

\keywords{   KubeEdge \and Edge Computing \and Internet of Things (IoT) \and Service Availability \and Lightweight Intrusion Detection \and Container Security.}
\end{abstract}
\section{Introduction}

The rapid expansion of distributed systems, edge computing, and IoT solutions is transforming modern Information Technology infrastructures \cite{b1}. With billions of connected devices generating continuous streams of data, local processing has become essential to reduce latency, optimize resource utilization, and ease network congestion.

In this context, KubeEdge an extension of Kubernetes tailored for edge computing \cite{b2} has emerged as a key technology. It enables the execution and management of containerized workloads directly on edge nodes, providing scalability and flexibility while reducing reliance on centralized cloud resources \cite{b3}.

However, the increasing reliance on containerization technologies such as Docker and orchestration platforms including Kubernetes and KubeEdge in critical domains (e.g., healthcare, transportation, and industrial automation) raises significant concerns regarding system security and resilience. These applications often require extremely high availability, frequently exceeding 99.999\% uptime \cite{b21}, to minimize downtime and ensure service continuity.

While KubeEdge leverages microservices and containerized workloads to enhance availability through fault isolation, rapid recovery, and lightweight orchestration \cite{b22}, these same architectural features introduce new security challenges. In particular, container images may be compromised through malicious code injection, outdated dependencies, or supply chain attacks \cite{b4,b5,b19}. For instance, the 2018 Docker Hub incident involved over five million downloads of compromised container images embedding cryptomining malware \cite{b10}.

These risks are further heightened in edge environments, where nodes typically operate under constrained computational resources and are often deployed in physically exposed or geographically distributed settings. This expanded attack surface increases the likelihood of both remote and physical intrusions. 

Moreover, security misconfigurations in pod policies, the use of unverified container images, and weaknesses in orchestration workflows can enable unauthorized code execution, privilege escalation, and even full node compromise \cite{b39,b93}. In addition, the continuous communication between cloud and edge nodes over public or semi-secured networks exposes systems to threats such as interception, hijacking, and Denial of Service (DoS) attacks, which can severely impact system performance and reliability.

In this work, we investigate how such vulnerabilities affect the availability and stability of edge services. While availability is typically defined as the proportion of time a system remains operational, we define stability as the system’s ability to maintain predictable behavior and recover autonomously during security incidents. Although stability is not formally quantified in this study, it is considered a key indicator of system resilience, observed through experimental behavior and discussed as a direction for future work.

Despite the existence of numerous Intrusion Detection Systems (IDS), most are designed for cloud-scale infrastructures and remain too resource-intensive for deployment in resource-constrained edge environments. This highlights the need for lightweight and adaptive security mechanisms specifically tailored to edge computing platforms.

Motivated by these challenges, this study evaluates the security exposure of IoT-related container images, simulates realistic attack scenarios such as denial-of-service and code injection, and assesses the effectiveness of a lightweight Intrusion Detection System in improving the resilience of KubeEdge-based deployments. To guide this investigation, we formulate the following research questions:

\begin{itemize}
    \item \textbf{RQ1:} What security vulnerabilities in IoT-related container images could facilitate attacks such as code injection, malicious pod deployment, or denial-of-service in KubeEdge-based environments?
    
    \item \textbf{RQ2:} How do these vulnerabilities affect the availability and stability of KubeEdge-based architectures?
    
    \item \textbf{RQ3:} To what extent can a lightweight Intrusion Detection System improve the availability and resilience of such environments?
\end{itemize}

The remainder of this paper is structured as follows. Section 2 introduces the background and foundational concepts related to Kubernetes and KubeEdge. Section 3 reviews the related work on security and availability in cloud–edge environments. Section 4 presents the proposed methodology, including the experimental setup, dataset construction, and the design of the intrusion detection mechanism. Section 5 reports and discusses the experimental results with respect to the research questions. Section 6 outlines the threats to validity. Finally, Section 7 concludes the paper and highlights future research directions.

\section{Background}

This section introduces the fundamental concepts underlying Kubernetes and KubeEdge, which are essential for understanding the context and technical foundations of this study.

\subsection{Kubernetes}

Kubernetes, introduced by Google in 2014, has become the de facto standard for orchestrating containerized applications in distributed and cloud-native environments \cite{b38}. Its architecture is designed to ensure scalability, reliability, and automated resource management across clusters\cite{b37}.

As illustrated in Figure 1, Kubernetes follows a control plane–worker node architecture. The control plane is responsible for managing the global state of the system and coordinating all cluster operations. It consists of several key components. The API Server acts as the central communication interface, exposing a RESTful API through which users and system components interact with the cluster. The Scheduler assigns Pods to worker nodes based on resource availability and scheduling policies. The Controller Manager continuously monitors the cluster state and enforces the desired configuration by handling tasks such as replication and failure recovery. The system state is persistently stored in etcd, a distributed key–value store that maintains configuration data and metadata.

The worker nodes are responsible for executing containerized workloads. Each node runs a kubelet, which ensures that containers are deployed and running according to the control plane’s instructions. The container runtime (e.g., Docker) manages container execution. Applications are encapsulated within Pods, the smallest deployable unit in Kubernetes, which may contain one or more containers sharing networking and storage resources. Additionally, kube-proxy manages network communication by routing traffic between services within the cluster.

This architecture enables Kubernetes to provide essential features such as automated deployment, scaling, and self-healing, forming the foundation of modern cloud-native systems.

\begin{figure}[h]
    \centering
    \includegraphics[width=0.9\linewidth]{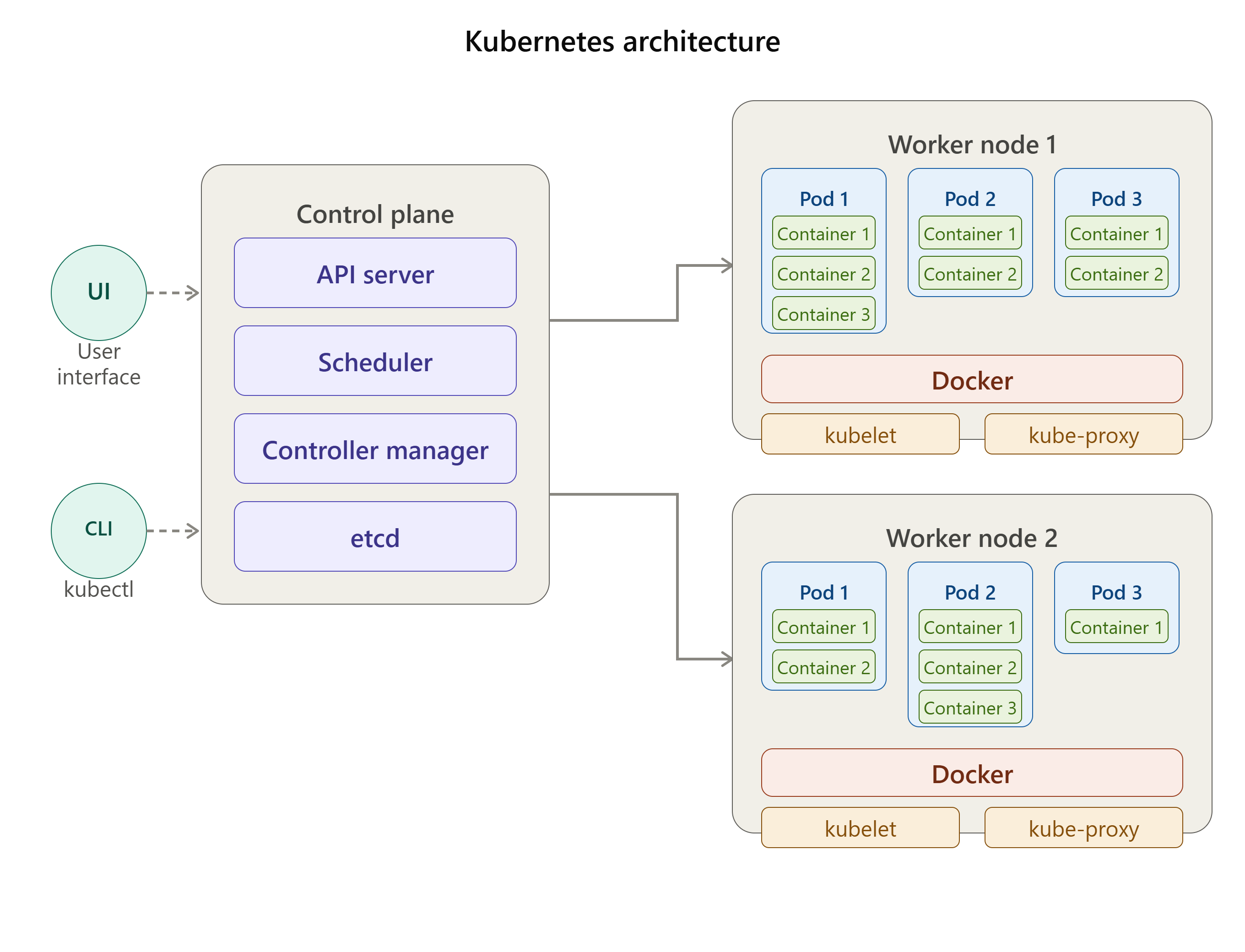}
    \caption{Kubernetes Architecture }
    \label{fig:kubernetes}
\end{figure}

\vspace{-8pt}

\subsection{KubeEdge}
KubeEdge extends Kubernetes capabilities to edge environments, which often operate under constrained resources. While this improves local autonomy and reduces latency, it also raises new security concerns \cite{ostif2022kubeedge}. By enabling seamless orchestration between cloud and edge nodes, KubeEdge allows applications to be deployed closer to data sources while maintaining centralized coordination.

As illustrated in Figure 2, the KubeEdge architecture is composed of two main subsystems: CloudCore and EdgeCore, which communicate through a bidirectional messaging infrastructure.

The CloudCore, deployed in the cloud, is responsible for global orchestration and cloud-to-edge coordination. It includes the EdgeController, which manages Pods and edge nodes, and the DeviceController, which handles IoT device lifecycle management. Communication between the cloud and edge layers is ensured by CloudHub, which facilitates synchronization of system state and metadata between CloudCore and EdgeCore.

The EdgeCore, deployed on edge nodes, is responsible for local workload execution and interaction with IoT devices. It includes several components. EdgeHub maintains communication with CloudHub and synchronizes cloud and edge states. Edged, the edge equivalent of Kubernetes kubelet, manages containerized applications locally. MetaManager handles local metadata storage and ensures continued operation during intermittent connectivity. DeviceTwin maintains the state of connected devices and synchronizes it with the cloud, enabling real-time monitoring and control. Additionally, EventBus enables communication with IoT devices through protocols such as MQTT, while ServiceBus supports service-level interactions with external applications.

Despite its advantages, the distributed and resource-constrained nature of KubeEdge introduces additional challenges. In particular, edge nodes are more exposed to security risks, including configuration errors, vulnerabilities in container images, and potential supply chain attacks. These issues can directly impact system availability and reliability, highlighting the need for lightweight and decentralized security mechanisms tailored to edge environments.

\begin{figure}[h]
    \centering
    \includegraphics[width=0.8\linewidth]{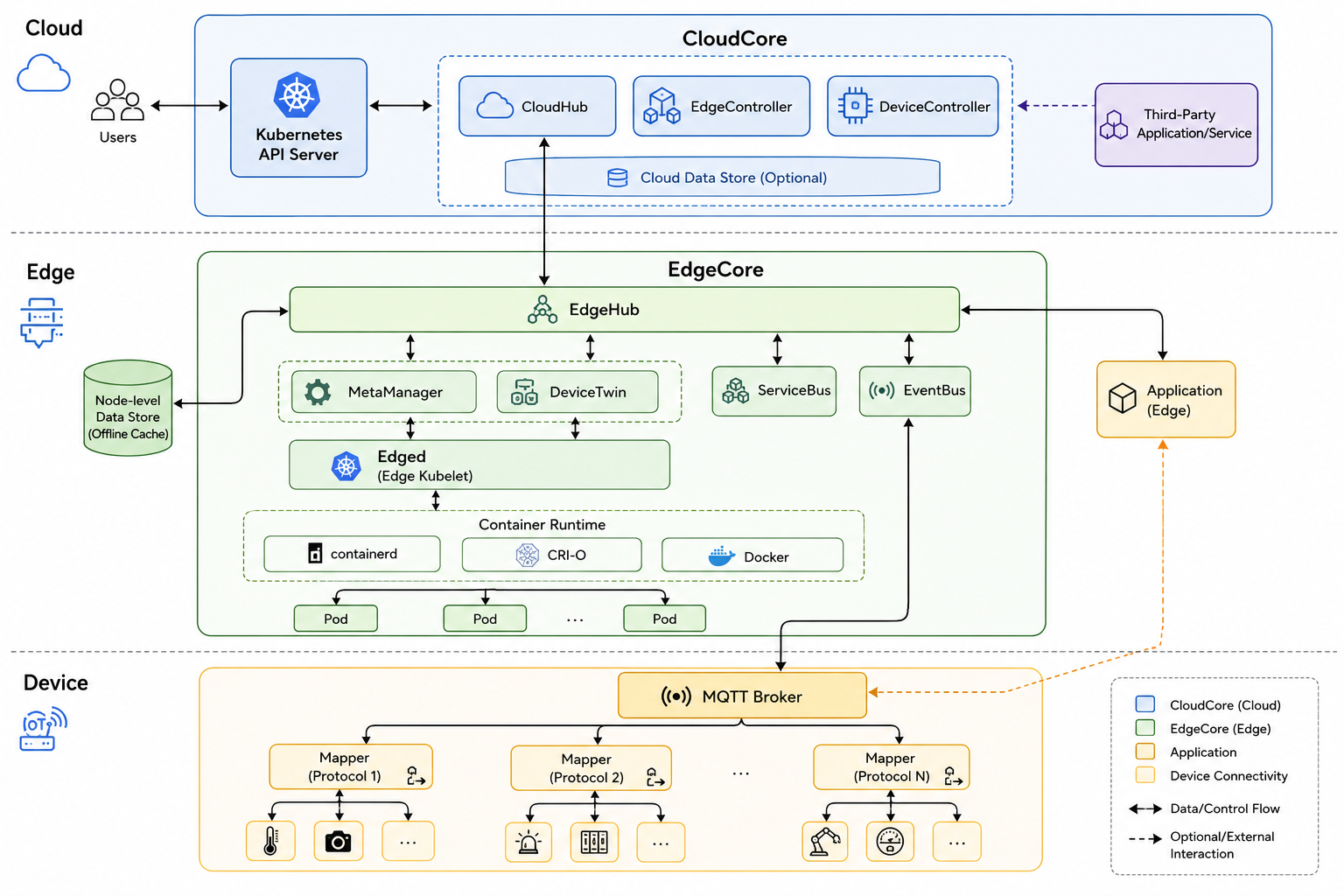}
    \caption{KubeEdge Architecture}
    \label{fig:kubernetes}
\end{figure}

\section{Related Work}

Kubernetes and its edge extensions, such as KubeEdge, have significantly advanced the deployment and orchestration of distributed applications across cloud–edge environments. However, despite these advances, ensuring both security and availability remains a major challenge, particularly in edge settings where resource constraints, decentralized execution, and increased exposure to attacks fundamentally alter system behavior. In such environments, misconfigurations, vulnerabilities in containerized workloads, and unstable connectivity can rapidly propagate, directly affecting system reliability.

Against this backdrop, several works have investigated security vulnerabilities in KubeEdge and edge computing systems. Korczynski et al.~\cite{ostif2022kubeedge} conducted a comprehensive security audit of KubeEdge, identifying a wide spectrum of threats, including internal misconfigurations, external attacks, and supply chain vulnerabilities. While their study provides valuable insights into the attack surface, it remains primarily diagnostic and does not explore practical mitigation strategies under runtime constraints. In a similar vein, Xiao et al.~\cite{b16} analyzed targeted attacks on distributed edge services, showing how vulnerabilities in containerized applications and orchestration frameworks can directly impact system availability. However, their work focuses mainly on attack characterization, leaving open the question of how such threats can be effectively detected and mitigated in real time.

To address security concerns in Kubernetes environments, prior research has proposed centralized enforcement mechanisms. For instance, Al-Obaidi et al.~\cite{b15} introduced an extension that enforces centralized validation of orchestration actions. Although this approach provides strong guarantees in terms of policy enforcement, it inherently introduces a single point of control, which may become both a bottleneck and a point of failure. More importantly, such centralized designs are poorly suited to edge environments, where network instability and decentralization are intrinsic characteristics.

Parallel to these efforts, a substantial body of work has focused on improving system availability and fault tolerance. Abdollahi Vayghan et al.~\cite{b18} proposed a Kubernetes controller that enhances the availability of elastic stateful microservices through automated failover and replication. While effective in reducing recovery time, their approach assumes the availability of sufficient computational resources and stable infrastructure, which limits its applicability in resource-constrained edge contexts.

Further insights are provided by empirical studies highlighting the limitations of Kubernetes in practice. Vayghan et al.~\cite{b35} demonstrated that, despite built-in self-healing mechanisms, Kubernetes does not inherently guarantee high availability. In particular, their results show that service outage durations can be significantly longer than expected in scenarios involving node or Pod failures. This observation underscores a critical gap between theoretical resilience guarantees and observed system behavior in real-world deployments.

In a follow-up study, Vayghan et al.~\cite{b36,b90} proposed a controller designed to improve the availability of stateful microservices. Their approach leverages state replication and dynamic request redirection, achieving substantial improvements in recovery time (ranging from 55\% to 99\%). Nevertheless, this solution introduces additional system complexity and overhead, and remains fundamentally reactive, as it operates after failures occur rather than preventing them.

Beyond Kubernetes-centric approaches, research in IoT-edge systems has explored alternative orchestration and monitoring strategies. Mouine and Saied~\cite{b17,b91} proposed an event-driven framework based on MQTT for distributed coordination across edge devices. While effective in enabling interoperability, their approach does not incorporate runtime security enforcement or intrusion detection capabilities. Similarly, Muralidharan et al.~\cite{b26,b92} focused on Kubernetes-based monitoring architectures for IoT infrastructures, emphasizing observability while leaving security concerns largely unaddressed.

Taken together, these works highlight several recurring limitations. First, many approaches rely on centralized architectures, which are inherently misaligned with the decentralized nature of edge environments. Second, existing solutions often assume stable connectivity and sufficient computational resources, limiting their applicability in real-world deployments. Third, and most importantly, the majority of existing techniques adopt a reactive paradigm, focusing on failure recovery (e.g., restart, replication, failover) rather than proactively detecting and mitigating anomalous behavior during execution.

These observations point to a fundamental gap in the literature: the absence of lightweight, decentralized, and runtime-aware security mechanisms capable of operating effectively under the constraints of edge computing.

To bridge this gap, this work introduces a fundamentally different detection paradigm tailored to edge-native environments. The proposed Recommended Intrusion Detection Rule Set (RIDRS) is designed to be lightweight, decentralized, and autonomously reactive, enabling security enforcement directly at the edge node without relying on centralized control.

This paradigm is grounded in three key principles. First, locality of defense ensures that detection and response mechanisms operate directly on edge nodes, thereby reducing latency and minimizing dependence on cloud connectivity. Second, security-by-design at the node level embeds detection capabilities within each node, enabling autonomous protection and enhancing overall system resilience. Third, runtime availability preservation enables the system to detect and mitigate anomalous behavior during execution (e.g., abnormal resource usage or suspicious network activity), thereby reducing mitigation time and preventing cascading failures.

While further large-scale evaluation is required to fully assess its effectiveness, this work provides a proof of concept demonstrating the feasibility of decentralized intrusion detection in KubeEdge-based IoT deployments. In this regard, RIDRS complements existing approaches by introducing an embedded, proactive, and edge-native defense layer, specifically designed to address the limitations of current orchestration-centric and reactive solutions.

\section{Methodology}

This section presents the experimental design adopted to investigate the impact of security vulnerabilities on the availability of KubeEdge-based systems. It describes the experimental setup, the dataset construction and vulnerability analysis process, the design of attack scenarios, the evaluation metrics, and the proposed lightweight intrusion detection mechanism.

\subsection{Experimental Setup}
To evaluate system behavior under realistic conditions, we deployed a KubeEdge-based experimental testbed that emulates a typical IoT-edge environment. The experimental KubeEdge cluster consists of three edge nodes: one primary node hosting the active workloads and two secondary nodes configured for failover. All nodes are implemented using Raspberry Pi 3 B+ devices running Ubuntu 22.04, Docker 20.10.14, and KubeEdge version 1.12.1. The cluster is orchestrated from a virtual machine acting as the CloudCore node, which is responsible for coordinating communication and workload management across edge nodes.

\begin{figure}[h!]
    \centering
    \includegraphics[width=0.8\linewidth]{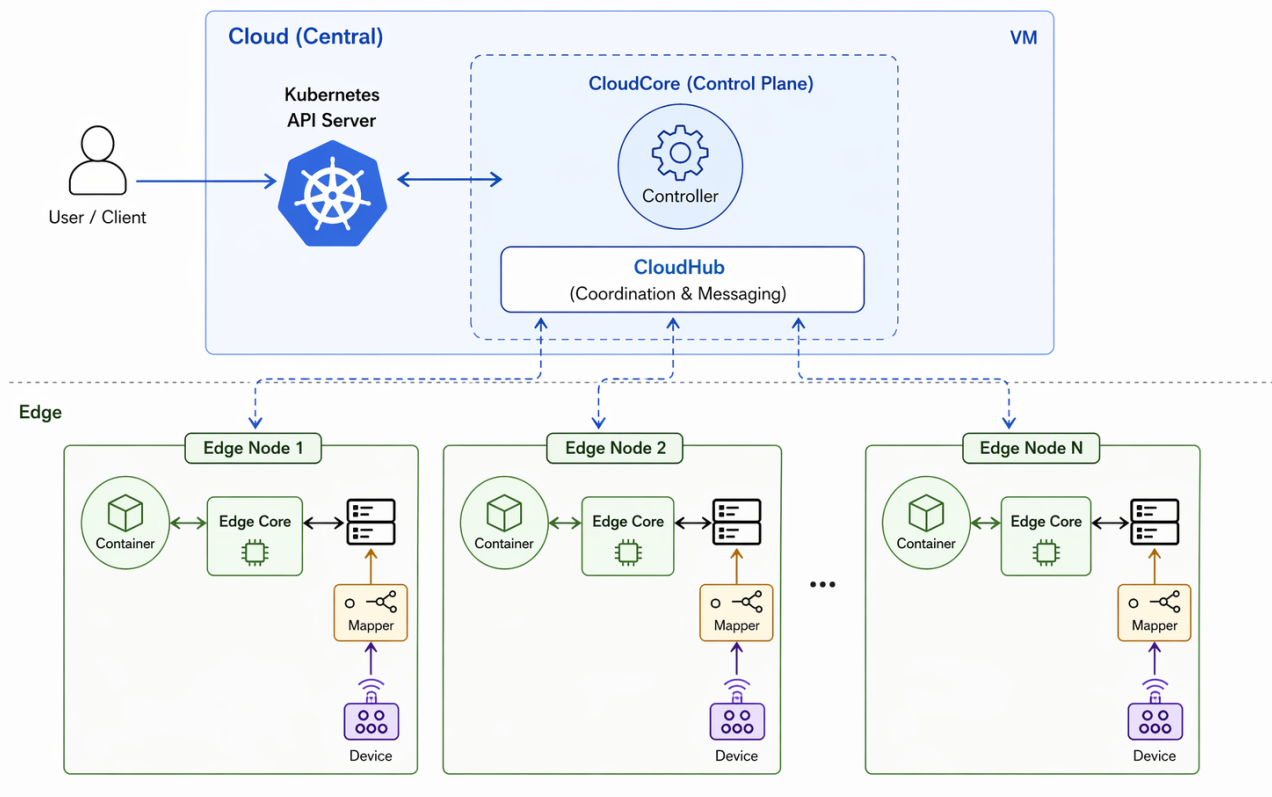}
    \caption{Deployment Architecture}
    \label{fig:Deployment}
\end{figure}

The deployment architecture, illustrated in Fig.~\ref{fig:Deployment}, reflects a cloud–edge configuration where distributed EdgeCore components execute containerized workloads and interact with IoT devices. The network configuration enables automatic pod rescheduling across nodes in the event of failures. This design emulates real-world edge deployments, where failover mechanisms play a critical role in maintaining service continuity under node failures or attack conditions.

To ensure accurate measurement of time-dependent metrics, synchronization between nodes is achieved using the Network Time Protocol (NTP) provided by Kubernetes.

For workload generation and system monitoring, we deployed a lightweight counter application. A container image embedding the counter is included in the pod template. Once a pod is deployed, a container instance is created from this image and continuously increments a counter, exposing outputs both through the console and a web interface. This setup enables real-time observation of system behavior and availability during experiments.

\subsection{Dataset Construction and Vulnerability Analysis (RQ1)}

Previous studies \cite{b9,b19} have shown that container images may contain critical vulnerabilities; however, limited work has quantitatively assessed the security quality of images specifically used in edge and IoT deployments. This gap motivates our first research question (RQ1), which aims to systematically assess the vulnerability exposure of container images tagged for use in KubeEdge-based and IoT-focused environments.

In this study, we adopted a three-step methodological approach to analyse container image vulnerabilities in edge computing and IoT contexts. The first phase involved collecting a total of 500 container images from public registries, including DockerHub and ArtifactHub. These images were selected based on the \textit{IoT} tag, with the objective of ensuring diversity by including both recent and older versions, thereby enabling a broader coverage of potential vulnerability profiles.

To ensure the relevance and accuracy of the dataset, a manual verification process was conducted for each image. This process involved examining image descriptions and functionalities to confirm their applicability to IoT use cases. The verified images cover a wide range of applications, including home automation platforms (e.g., Home Assistant), IoT gateways (e.g., Eclipse Kura), sensor data collection and processing systems, as well as industrial equipment monitoring applications.

The second phase focused on conducting a security audit of the verified images using the vulnerability detection tool Trivy \cite{b23}. Trivy was used to identify known vulnerabilities (CVEs) and classify them according to their severity levels, ranging from LOW to CRITICAL. The analysis also considers vulnerabilities introduced through embedded software dependencies, including outdated libraries and misconfigurations that may expose the system to exploitable weaknesses.

\subsection{Attack Scenarios and Evaluation Metrics (RQ2)}

The resilience of a KubeEdge-based system is closely tied to its ability to detect and mitigate security threats that affect node behavior and service delivery. Due to the decentralized nature of edge computing and the resource constraints of edge nodes, traditional security mechanisms are often insufficient. As a result, even minor vulnerabilities, when exploited, may significantly impact system availability.

To evaluate the practical impact of such threats, we designed a set of attack scenarios derived from known vulnerabilities in container images. These scenarios aim to reproduce plausible exploitation conditions affecting different layers of the system. The study focuses on the following attack types:

\begin{itemize}
    \item \textbf{Denial-of-Service (DoS) Attacks:}  
    DoS attacks were simulated by generating high-volume traffic toward edge nodes using tools such as \texttt{hping3} and \texttt{tcpdump}. These attacks aim to exhaust system resources, including CPU, memory, and network bandwidth, in order to observe the resulting degradation in service availability.

    \item \textbf{Malicious Container Injection:}  
    This scenario involves deploying compromised container images within the KubeEdge cluster to evaluate the propagation of malicious workloads and their impact on cluster stability. The experiment leverages weak authentication mechanisms and misconfigured Kubernetes Role-Based Access Control (RBAC) policies to escalate privileges.

    \item \textbf{Code Injection Attacks:}  
    Code injection attacks were implemented by exploiting vulnerabilities in containerized applications and runtime environments. In this scenario, arbitrary payloads are injected to trigger abnormal system behavior, enabling the evaluation of remote code execution effects within the KubeEdge framework.
\end{itemize}

Each attack scenario was monitored using real-time logging, resource usage metrics (CPU and memory), and service availability tracking. This monitoring setup enables the observation of system behavior throughout the execution of each attack.

To systematically evaluate the impact of these attacks, we define a set of temporal metrics that describe the lifecycle of a failure. The introduction time corresponds to the moment when the attack begins to affect system behavior. The KubeEdge reaction time represents the delay between the occurrence of anomalies and their detection by system components such as Edged and kubelet.

The intervention time corresponds to the duration required to initiate recovery actions, such as restarting containers or redeploying pods. The recovery time represents the time needed for the system to return to a stable operational state. Finally, the outage time is defined as the total duration during which the service remains unavailable.

These metrics provide a structured framework for evaluating system resilience and enable consistent comparison across different attack scenarios.

\subsection{Lightweight Intrusion Detection Mechanism (RQ3)}

To address the unique challenges of securing edge computing environments, we designed a lightweight, rule-based detection mechanism specifically tailored for KubeEdge architectures. Rather than implementing a full Intrusion Detection System (IDS), we propose a \textbf{Recommended Intrusion Detection Rule Set (RIDRS)} composed of focused behavioral rules. This RIDRS is designed to operate with minimal CPU and memory overhead, making it suitable for deployment on resource-constrained edge nodes such as Raspberry Pi devices.

The detection mechanism relies on predefined behavioral rules targeting attack patterns commonly observed in IoT and edge environments. These rules can be integrated into existing open-source IDS tools such as \textit{Falco}, or deployed independently as part of lightweight monitoring agents. When a rule is triggered, the system logs the event and optionally initiates mitigation actions.

\paragraph{Rule 1: Excessive Memory Consumption}

This rule monitors the memory usage of active processes to detect potential memory exhaustion attacks (e.g., crypto-mining or memory leaks). Instead of using a fixed threshold, a \textbf{dynamic threshold} is defined as follows:

\[
\text{Threshold} = \text{Average Normal Usage} + \alpha \times \text{Total RAM}
\]

where:
\begin{itemize}
    \item \textbf{Average Normal Usage} represents the typical memory consumption observed during a baseline period,
    \item \textbf{Total RAM} is the total available memory on the device,
    \item \textbf{$\alpha$} is an adjustable margin (set to 15\% in our experiments).
\end{itemize}

\textit{Detection logic:}
\begin{itemize}
    \item Periodically sample memory usage of all active processes using \texttt{ps aux}.
    \item Compute the dynamic threshold after a warm-up period.
    \item If a process exceeds the threshold, log the event and optionally terminate the process.
\end{itemize}

\paragraph{Rule 2: Excessive CPU Usage}

This rule detects processes that continuously consume high CPU resources (e.g., above 80\%), which may indicate denial-of-service behavior or unauthorized computation.

\textit{Detection logic:}
\begin{itemize}
    \item Continuously monitor CPU usage using tools such as top or ps.
    \item If a process exceeds the defined threshold for a sustained period, generate an alert and optionally terminate the process.
\end{itemize}

\paragraph{Rule 3: ICMP Flood Detection (HPING3 Attack)}

This rule identifies high volumes of ICMP echo-request packets, typically associated with ping flood or HPING3-based DoS attacks. It inspects network traffic and mitigates such behavior at the firewall level.

\textit{Detection logic:}
\begin{itemize}
    \item Capture ICMP traffic using
    
    \texttt{tcpdump 'icmp[icmptype] == 8'}.
    \item If flood behavior is detected, block incoming ICMP echo-requests using:
    
    \texttt{iptables -A INPUT -p icmp --icmp-type echo-request -j DROP}.
\end{itemize}

\paragraph{Mitigation Strategy}

Upon detecting a rule violation, the system performs the following actions:
\begin{itemize}
    \item Log the suspicious activity for auditing and analysis,
    \item Optionally terminate offending processes to prevent further impact,
    \item Dynamically update firewall rules to block malicious traffic.
\end{itemize}

This modular and rule-based design enables easy extension with additional detection rules or integration with advanced threat intelligence mechanisms. Despite its simplicity, RIDRS provides effective detection and mitigation capabilities while maintaining a low computational footprint, making it well-suited for edge environments.

\section{Results and Discussion}

This section presents and discusses the results obtained from the experimental evaluation. The findings are organized according to the three research questions, focusing on vulnerability exposure, the impact of attacks on system availability, and the effectiveness of the proposed detection mechanism.

\subsection{Vulnerability Distribution and Risk Exposure (RQ1)}

Assessing the vulnerability landscape of container images is a key step toward understanding security risks in edge computing environments. To this end, we analyze the vulnerability landscape of IoT container images to characterize both their prevalence and severity.

\begin{figure}[H]
    \centering
    \includegraphics[width=0.7\linewidth]{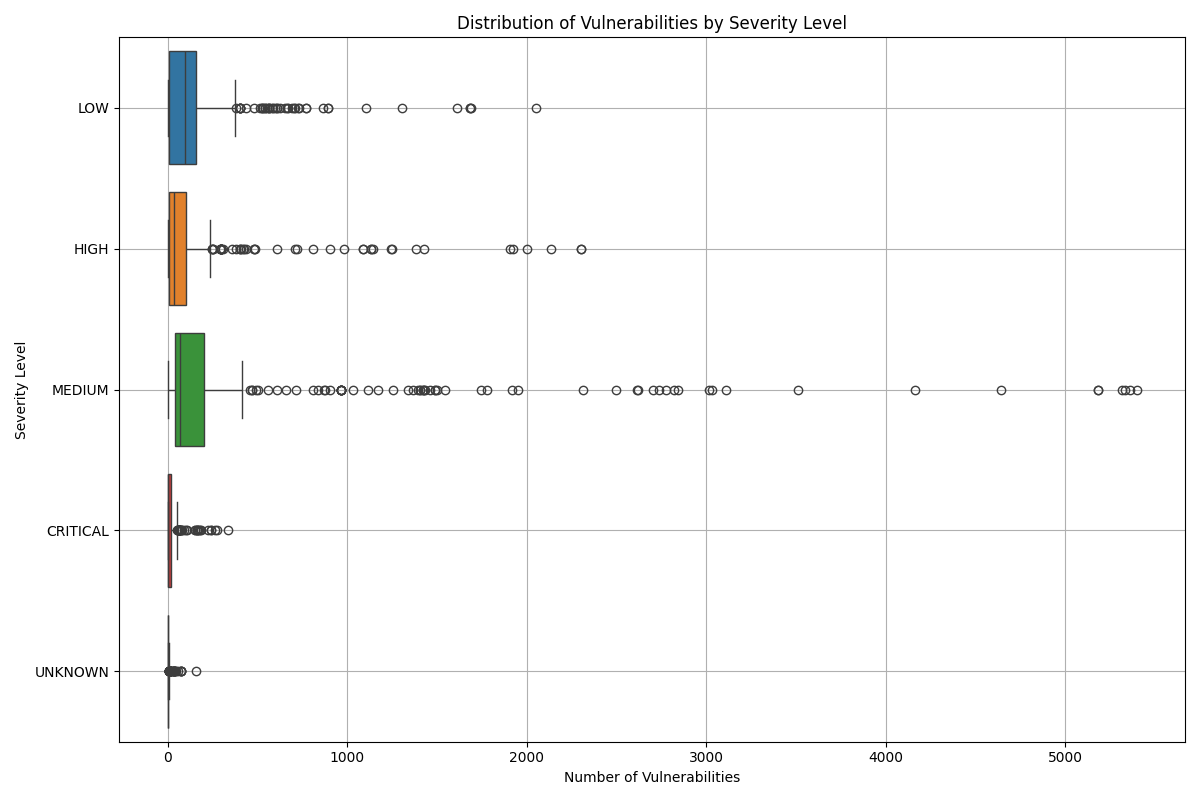}
    \caption{Boxplot of vulnerability distribution per image by severity}
    \label{fig:vuln_boxplot}
\end{figure}

Figure~\ref{fig:vuln_boxplot} presents the distribution of vulnerability counts per container image across different severity levels (Critical, High, Medium, Low, and Unknown). This visualization reveals a high degree of variability in the security posture of the analyzed images, indicating that vulnerability exposure is not uniformly distributed.

A first key observation is the pervasive nature of vulnerabilities across the dataset. More than 95\% of the analyzed images contain vulnerabilities spanning multiple severity levels, including HIGH and CRITICAL categories. This widespread exposure suggests that security issues are not confined to a small subset of poorly maintained images but are instead deeply embedded within the IoT container ecosystem.

A deeper analysis shows that the total number of vulnerabilities is not a reliable indicator of actual risk. Several images with fewer than ten vulnerabilities still include HIGH or CRITICAL issues. In edge environments such as KubeEdge, even a single critical vulnerability may be sufficient to compromise system availability or enable privilege escalation. This finding highlights the limitations of vulnerability-count-based assessments and underscores the importance of severity-aware evaluation.

At the same time, the distribution exhibits a strong imbalance, with a small subset of images concentrating a large proportion of vulnerabilities. Extreme cases illustrate this phenomenon. For example, the image marcu5fen1x/iot contains up to 273 critical vulnerabilities, indicating a particularly high-risk artifact. High-severity vulnerabilities are more broadly distributed, with notable outliers such as nehavadnere/iot (1909 vulnerabilities) and liduck/iot (2133 vulnerabilities). Medium-severity vulnerabilities are the most prevalent and display the widest dispersion, with some images exceeding 5000 occurrences, resulting in a heavy-tailed distribution. Low-severity vulnerabilities are also frequent but less dispersed, while UNKNOWN vulnerabilities remain relatively limited and appear in isolated cases such as berryjamccl/iot, suggesting incomplete metadata or recently disclosed issues.

\begin{figure}[H]
    \centering
    \includegraphics[width=1.02\linewidth]{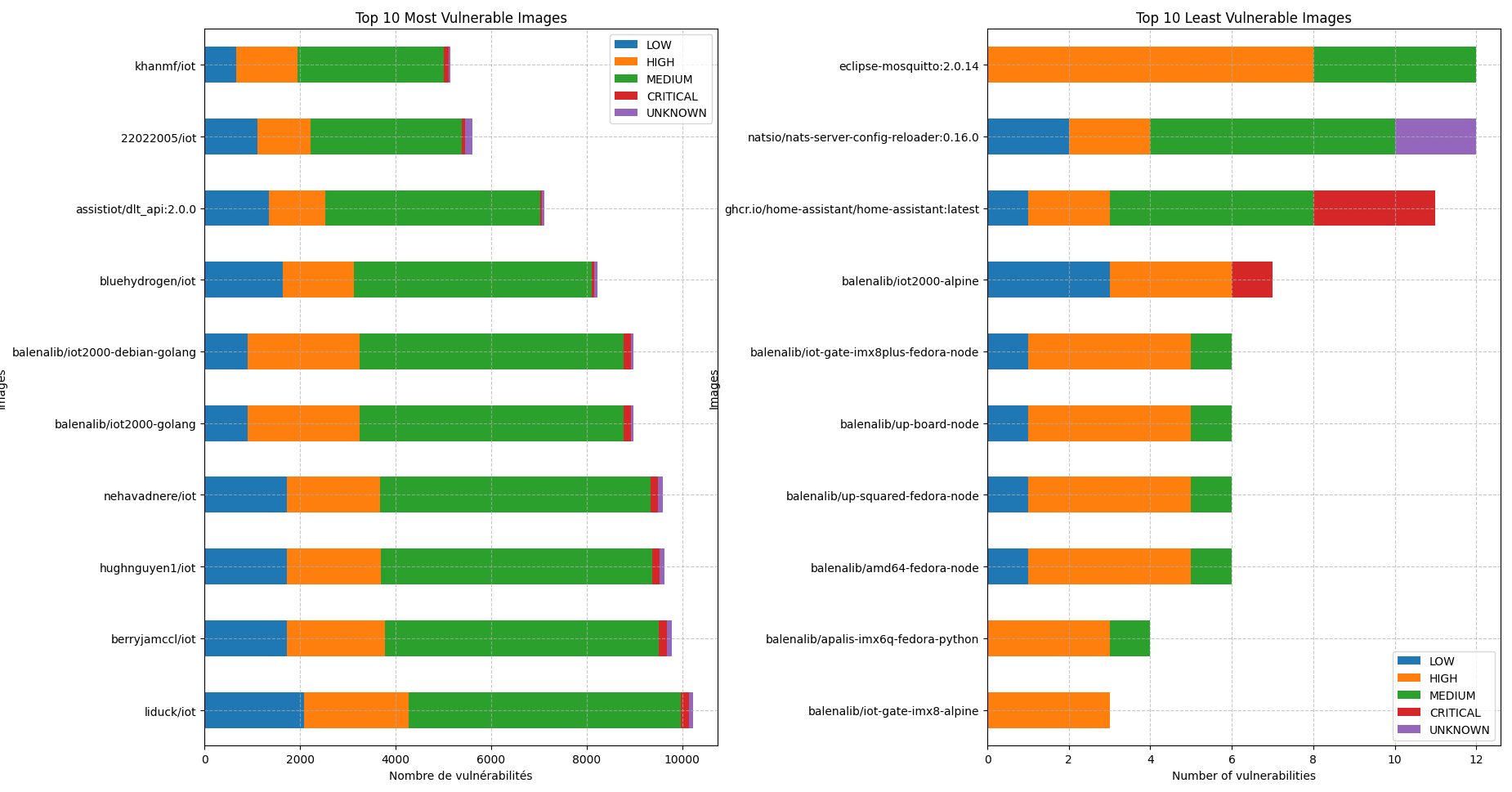}
    \caption{Top 10 Most and Least Vulnerable Images}
    \label{fig:top10_vuln_images}
\end{figure}

Figure~\ref{fig:top10_vuln_images} further reinforces these observations by contrasting the ten most and ten least vulnerable container images. The most affected images, including liduck/iot, nehavadnere/iot, and bluehydrogen/iot, accumulate several thousand vulnerabilities, predominantly in the MEDIUM and HIGH categories. In contrast, the least vulnerable images, such asbalenalib/iot-gate-imx8-alpine and balenalib/apalis-imx6q-fedora-python, contain fewer than ten vulnerabilities, reflecting stronger maintenance and security practices. However, even among these relatively secure images, some still exhibit CRITICAL or UNKNOWN vulnerabilities, as observed in ghcr.io/home-assistant/home-assistant.

From a global perspective, 478 images contain at least one HIGH vulnerability, 410 contain CRITICAL vulnerabilities, 477 include MEDIUM vulnerabilities, 471 include LOW vulnerabilities, and 194 contain UNKNOWN vulnerabilities. These results highlight a broad and overlapping exposure across severity levels, reinforcing the notion that vulnerability presence is widespread rather than exceptional.

Overall, the findings reveal a dual risk structure: while vulnerability exposure is pervasive across nearly all images, a relatively small subset of images accounts for a disproportionately large share of vulnerabilities. This combination of widespread exposure and concentrated risk underscores the necessity of systematic vulnerability assessment, severity-aware filtering, and automated validation mechanisms prior to deploying container images in resource-constrained edge environments such as KubeEdge.

\subsection{Impact of Security Attacks on System Availability (RQ2)}

To assess the impact of security vulnerabilities on system availability, we conducted controlled experiments simulating three attack scenarios: code injection, malicious pod deployment, and denial-of-service (DoS) attacks. Each scenario targets different layers of the KubeEdge architecture and allows us to observe system behavior under distinct failure conditions.

\begin{figure}[htbp]
    \centering
    \includegraphics[width=0.7\linewidth]{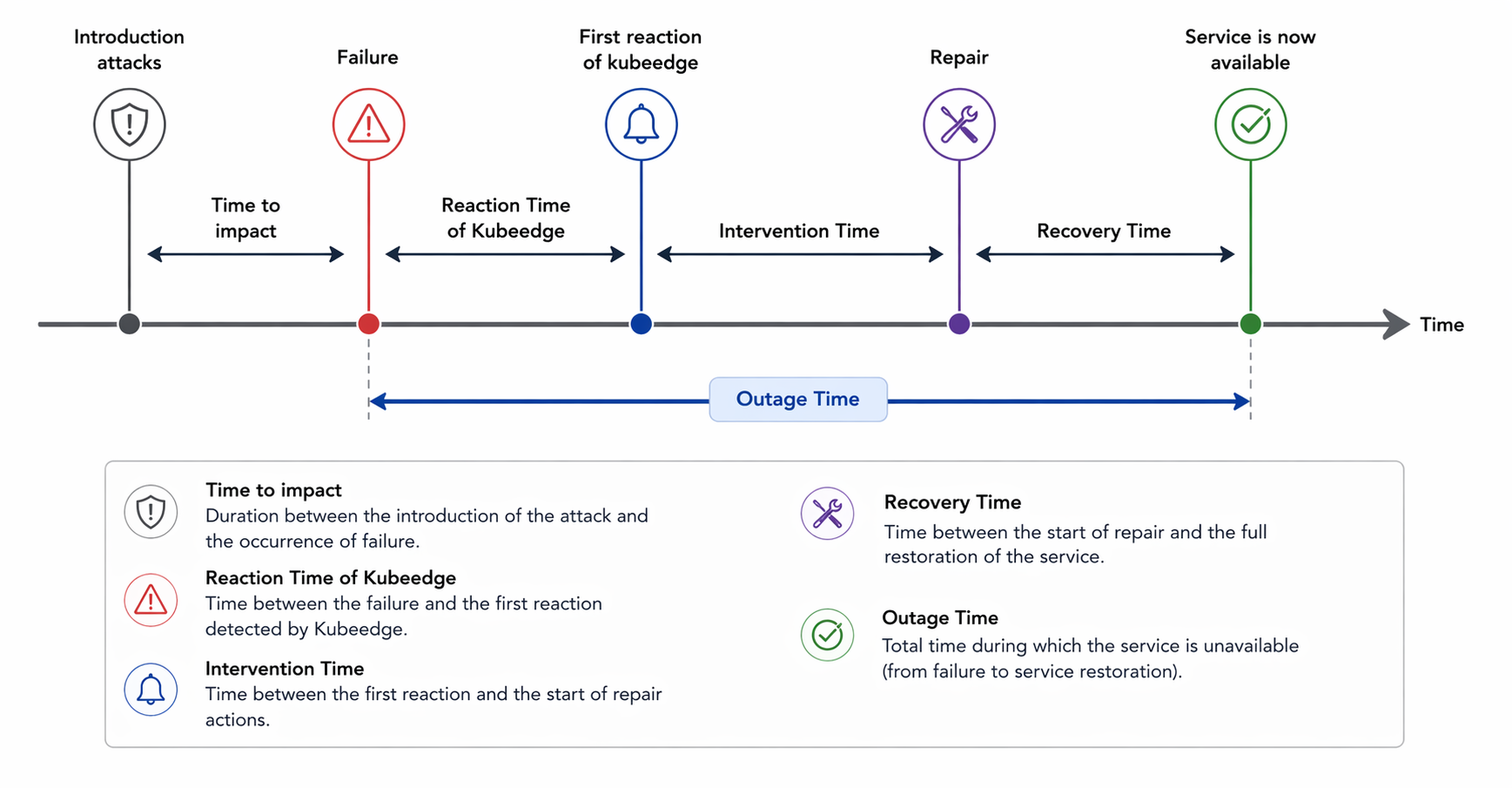}
    \caption{Availability metrics related to security issues}
    \label{fig:availabilitys}
\end{figure}

To systematically evaluate these impacts, we define a set of temporal metrics capturing the lifecycle of a failure. These include \textit{time to impact}, \textit{KubeEdge reaction time}, \textit{intervention time}, \textit{recovery time}, and \textit{outage time}. Figure~\ref{fig:availabilitys} illustrates the relationships between these metrics.

The experimental results, averaged over ten executions per scenario, are summarized in Table~\ref{tab:resilience1}.

\begin{table}[htbp]
\centering
\caption{Experimental metrics measuring the impact of our IDS on service availability in seconds}
\label{tab:resilience1}
\small
\renewcommand{\arraystretch}{1}
\begin{tabular}{|l|c|c|c|}
    \hline
    \textbf{Trigger} & \makecell{\textbf{Code}\\\textbf{Injection}} & \makecell{\textbf{Malicious} \\ \textbf{Pod}} & \makecell{\textbf{DoS}\\\textbf{Attack}} \\
    \hline
    \textbf{Time to Impact} & 2.12 s & 40.860 s & 17.103 s \\
    \hline
    \textbf{KubeEdge Reaction Time} & 9.914 s & 46.617 s & 24.583 s \\
    \hline
    \textbf{Intervention Time} & 9.985 s & $\infty$ & $\infty$ \\
    \hline
    \textbf{Recovery Time} & 1.278 s & $\infty$ & $\infty$ \\
    \hline
    \textbf{Outage Time} & 21.177 s & $\infty$ & $\infty$ \\
    \hline
\end{tabular}
\end{table}

\paragraph{Code Injection Scenario}

In the code injection scenario, the system exhibits the shortest service disruption, with an outage time of 21.17 seconds. The anomaly is detected within 9.91 seconds, followed by rapid intervention and recovery. Once the injected code disrupts the application, the operating system terminates the affected processes. The \texttt{Edged} component detects the failure and notifies the \texttt{kubelet}, which triggers the redeployment of a clean pod. This behavior demonstrates that KubeEdge can effectively handle localized failures affecting individual pods.

\paragraph{Malicious Pod Scenario}

In contrast, the malicious pod scenario leads to severe and persistent system degradation. The attack results in progressive resource exhaustion, leading to node saturation after approximately 40.86 seconds. Although KubeEdge detects the anomaly after 46.61 seconds, the recovery process fails. The failover mechanism reschedules the malicious pod onto another node, replicating the attack and causing cascading failures across the cluster. Consequently, both intervention and recovery times are effectively infinite, and the system remains unavailable until the attack is manually stopped.

\paragraph{Denial-of-Service (DoS) Scenario}

A similar limitation is observed in the DoS scenario. The attack saturates system resources and disrupts network communication, leading to service failure after 17.10 seconds. KubeEdge detects anomalies after 24.58 seconds, when connection errors are identified. However, no automatic recovery occurs, and service availability is restored only after the attack is externally mitigated.

These results highlight several key limitations of KubeEdge. First, while the system effectively handles localized failures, it struggles to recover from attacks that impact system-wide resources. Second, the automatic failover mechanism may become counterproductive in adversarial conditions, as it can propagate malicious workloads across nodes and amplify the impact of attacks.

Finally, these findings emphasize the need for proactive detection mechanisms capable of identifying threats before resources are fully exhausted. Such mechanisms are essential to prevent cascading failures and improve the resilience of edge-native systems.

Moreover, most existing intrusion detection solutions are designed for cloud environments and are not well suited for deployment on resource-constrained edge devices. While tools such as \textit{Falco} provide container-level monitoring, they are not specifically adapted to the decentralized nature of KubeEdge. This motivates the need for lightweight and edge-aware detection approaches, as explored in the next section.

\subsection{Impact of RIDRS on Availability (RQ3)}

To evaluate the effectiveness of the proposed lightweight intrusion detection mechanism, we repeated the attack scenarios under the same experimental conditions with the integration of the Recommended Intrusion Detection Rule Set (RIDRS). The results demonstrate a significant improvement in system availability and resilience for internal attack scenarios.

\begin{figure}[!htbp]
    \centering
    \includegraphics[width=0.8\linewidth]{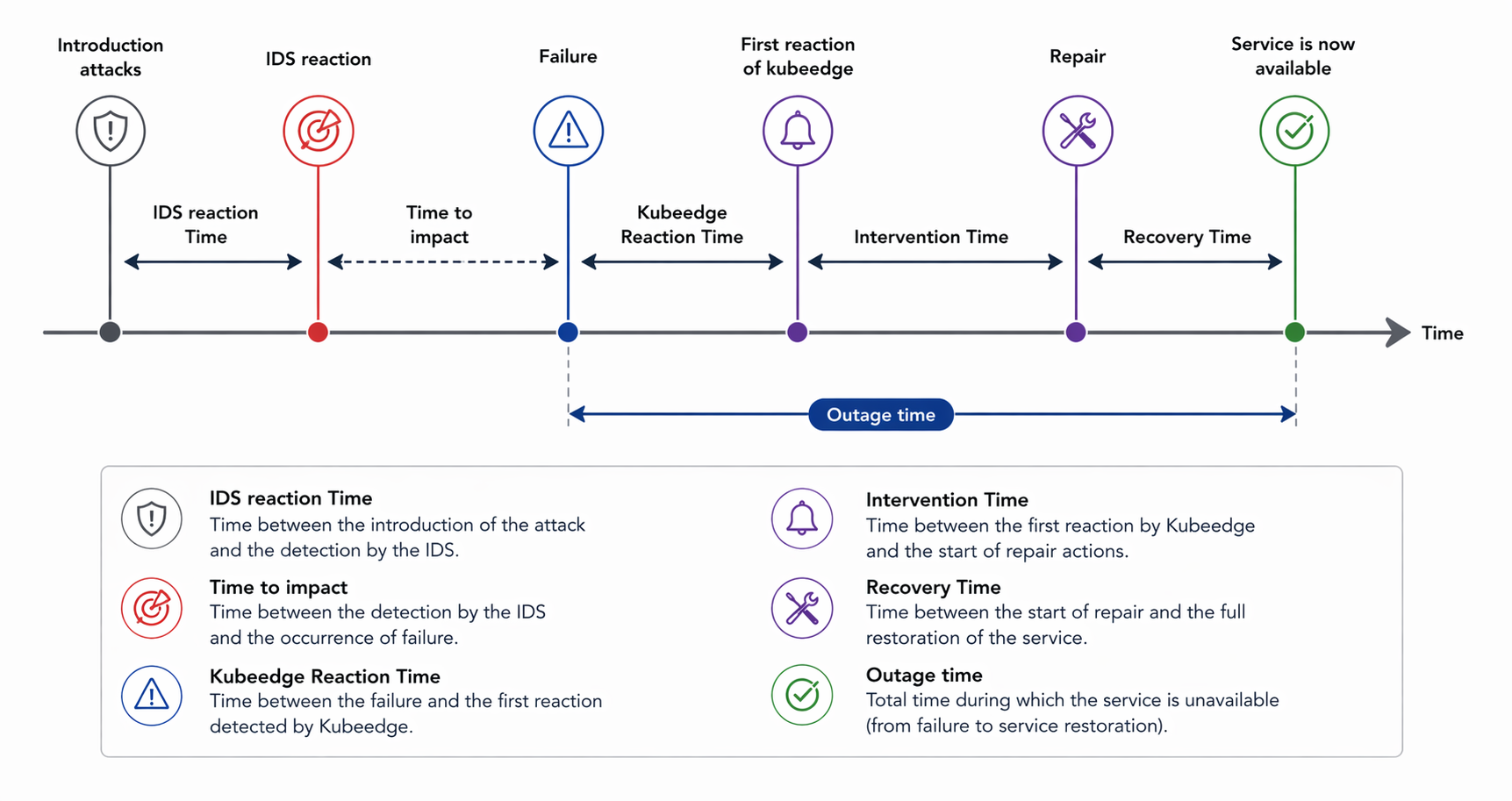}
    \caption{Availability metrics after introducing the proposed IDS}
    \label{fig:availability-ids}
\end{figure}

Figure~\ref{fig:availability-ids} illustrates the updated availability model after introducing the IDS. In addition to the previously defined metrics, we introduce a new metric, \textit{IDS reaction time}, which corresponds to the time interval between the introduction of an attack and its detection by the IDS.

The quantitative results, averaged over ten executions, are summarized in Table~\ref{tab:resilience}. These results highlight the effectiveness of RIDRS in detecting and mitigating internal threats before they impact system availability.

\begin{table}[h]
\centering
\caption{Experimental metrics measuring the impact of RIDRS on service availability (in seconds)}
\label{tab:resilience}
\small
\renewcommand{\arraystretch}{1}
\begin{tabular}{|l|c|c|c|}
    \hline
    \textbf{Trigger} & \makecell{\textbf{Code}\\\textbf{Injection}} & \makecell{\textbf{Malicious}\\\textbf{Pod}} & \makecell{\textbf{DoS}\\\textbf{Attack}} \\
    \hline
    \textbf{IDS Reaction Time} & 1.33 s & 30.12 s & 9.1 s \\
    \hline
    \textbf{Time to Impact} & 0 s & 0 s & 17.103 s \\
    \hline
    \textbf{KubeEdge Reaction Time} & 0 s & 0 s & 24.583 s \\
    \hline
    \textbf{Intervention Time} & 0 s & 0 s & $\infty$ \\
    \hline
    \textbf{Recovery Time} & 0 s & 0 s & $\infty$ \\
    \hline
    \textbf{Outage Time} & 0 s & 0 s & $\infty$ \\
    \hline
\end{tabular}
\end{table}

\paragraph{Analysis of Internal Attacks}

The integration of RIDRS eliminates service disruption for both code injection and malicious pod scenarios. As shown in Table~\ref{tab:resilience}, the IDS detects and neutralizes these attacks before any observable system degradation occurs, resulting in zero time to impact and zero outage time. In particular, code injection attacks are detected within 1.33 seconds, while malicious pod activity is identified within 30.12 seconds.

These results indicate a clear shift from a reactive recovery model to a proactive detection paradigm. Unlike the baseline scenario, where KubeEdge reacts only after failures occur, RIDRS enables early intervention, preventing resource exhaustion and avoiding cascading failures across nodes.

\paragraph{Impact on DoS Attacks}

In contrast, the DoS scenario remains largely unaffected by the proposed approach. Although the IDS detects abnormal traffic patterns within 9.1 seconds, the attack continues to saturate network resources, leading to persistent service unavailability. As reflected in Table~\ref{tab:resilience}, both intervention and recovery times remain effectively infinite.

This limitation highlights that node-level detection mechanisms alone are insufficient to mitigate large-scale network-based attacks, which affect the entire communication layer simultaneously.

Overall, the results demonstrate that RIDRS significantly enhances the resilience of KubeEdge-based systems against internal threats by enabling early detection and mitigation. The reduction of outage time to zero in two out of three scenarios represents a substantial improvement over the default behavior of KubeEdge.

Compared to existing solutions such as \textit{Falco}~\cite{b31} and \textit{Suricata}~\cite{b32}, which are primarily designed for cloud-scale environments, RIDRS is specifically tailored for resource-constrained edge devices, offering a lightweight alternative with minimal computational overhead.

However, the persistence of DoS-related failures indicates that additional network-level defense mechanisms are required to achieve comprehensive protection. This suggests that effective security in edge environments should combine lightweight local detection with complementary global mitigation strategies.

\textbf{Note:} The value $\infty$ is used to represent scenarios in which the system remains unavailable due to continuous attack conditions and the absence of automatic recovery mechanisms.

\section{Threats to Validity}

As with any empirical study, certain factors may influence the interpretation of the results. While the experimental design relies on controlled and repeatable scenarios, minor variations in system behavior or measurement precision may affect the reported timing metrics. The evaluation is based on well-established availability metrics (e.g., time to impact, reaction time, and outage time), which provide a consistent and interpretable framework, although they may not capture all aspects of long-term system dynamics. The experimental setup, built on a resource-constrained KubeEdge testbed using Raspberry Pi devices, is intentionally designed to reflect realistic edge conditions; however, results may vary in larger or more heterogeneous deployments. Similarly, the dataset of 500 IoT container images, selected and manually validated from public registries, provides a diverse and representative sample, although it does not exhaustively cover all possible deployment scenarios. Finally, the study focuses on a set of common and practically relevant attack scenarios, and while the proposed rule-based RIDRS mechanism demonstrates strong effectiveness against these threats, additional evaluation against more complex or adaptive attack strategies could further extend its applicability. Overall, these considerations do not undermine the validity of the findings but rather define the scope within which the results can be interpreted.

\section{Future Work}

Building on the results obtained with RIDRS, several directions can further extend this work.
First, integrating RIDRS with network-level mechanisms would enable a unified defense against large-scale attacks combining local detection and global mitigation.
Second, incorporating adaptive techniques such as automated rule refinement would enhance detection capabilities while preserving the lightweight nature of the approach.
Third, extending the evaluation to larger and more heterogeneous edge environments would further validate the robustness and generalizability of RIDRS.
Finally, broadening the scope beyond availability by incorporating additional resilience indicators and more advanced threat scenarios would strengthen the overall security guarantees of edge-native systems.

\section{Conclusion}

This paper studied the impact of security vulnerabilities on the availability of KubeEdge-based edge systems. Our analysis shows that vulnerabilities in IoT container images are both widespread and unevenly distributed, with critical risks present even in seemingly low-risk images. Experimental results demonstrate that while KubeEdge can handle localized failures, it remains vulnerable to attacks that affect shared resources or propagate across nodes, where failover mechanisms may amplify the impact.

To address these limitations, we proposed RIDRS, a lightweight rule-based intrusion detection mechanism tailored for resource-constrained edge environments. The results show that RIDRS significantly improves system resilience by enabling early detection and mitigation of internal attacks, effectively eliminating service disruption in multiple scenarios.

However, large-scale DoS attacks remain challenging, indicating that node-level detection must be complemented by network-level defenses. Overall, this work highlights the importance of proactive, lightweight, and decentralized security mechanisms for ensuring availability in edge computing systems.

%
%
%
%

\end{document}